\documentclass{ws-ijmpcs}
\usepackage[utf8]{inputenc}
\usepackage[english]{babel}
\usepackage[superscript]{cite}

\begin{document}

\markboth{Emanuele R. Nocera}
{Achievements and open issues in polarized PDF determinations}

%
\catchline{}{}{}{}{}
%

\title{Achievements and open issues\\ 
in the determination of polarized parton distribution functions}

\author{Emanuele R. Nocera}

\address{Dipartimento di Fisica, Universit\`{a} di Genova, Via Dodecaneso, 33\\
I-16146 Genova, Italy\\
emanuele.nocera@edu.unige.it}

\maketitle

\begin{history}
\received{Day Month Year}
\revised{Day Month Year}
\published{Day Month Year}
\end{history}

\begin{abstract}

I review the current status of the determination of helicity-dependent, or 
polarized, parton distribution functions from a comprehensive analysis of 
experimental data in perturbative quantum chromodynamics. I illustrate the 
latest achievements driven by new measurements in polarized proton-proton 
collisions at the Relativistic Heavy Ion Collider, namely the first evidence 
of a sizable polarized light sea quark asymmetry and of a positive polarized 
gluon distribution in the proton. I discuss which are the open issues in the 
determination of polarized distributions, and how these may be addressed in 
the future by ongoing, planned and proposed experimental programs.

\end{abstract}

\ccode{PACS numbers: 13.88.+e, 13.85.Ni}

\vspace{0.3cm}
\noindent
Understanding how the nucleon spin is built up from the spin of quarks and 
gluons - and their orbital angular momentum - is one of the most challenging 
goals in hadron physics~\cite{Bass:2007zzb}. For instance, according to the 
Jaffe and Manohar helicity sum rule~\cite{Jaffe:1989jz}
\begin{equation}
\frac{1}{2} 
=
\frac{1}{2}\Delta\Sigma(\mu^2) + \Delta G(\mu^2)
+\mathcal{L}_q(\mu^2) + \mathcal{L}_g(\mu^2)
\,\mbox{,}
\label{eq:helicitysumrule}
\end{equation}
the one-half proton spin can be explicitly decomposed into contributions from
quark and gluon spin, $\Delta\Sigma$ and $\Delta G$, and from quark and gluon
orbital angular momentum, $\mathcal{L}_q$ and $\mathcal{L}_g$.  Whether each 
of these terms allows for a unique field-theoretic 
definition~\cite{Collins:2011zzd} in Quantum Chromodynamics 
(QCD) - and is possibly a gauge-invariant, physically
meaningful, measurable quantity - has raised a major controversy in the last 
years, partly clarified only very recently~\cite{Leader:2013jra}.

Usually, Eq.~\eqref{eq:helicitysumrule} is
probed by measuring spin asymmetries, {\it i.e.} differences of cross sections 
with opposite polarizations of initial-state particles, that arise in a large 
variety of high-energy, large-momentum-transfer processes~\cite{Aidala:2012mv}.
Following factorization~\cite{Collins:1989gx}, 
these are described as a convolution between a short-distance
part (that contains information on the hard interactions of partons 
in the form of process-dependent kernels) and a long-distance part 
(that contains information on the spin structure of the nucleon in the 
form of universal parton distributions). The former can be computed in 
perturbative QCD; the latter should be determined from experimental data;
both depend on the factorization scheme and scale $\mu$.  

In this write-up, I review recent progress in the determination of 
helicity-dependent, or longitudinally polarized, Parton Distributions
Functions (PDFs) 
\begin{equation}
\Delta f(x,\mu^2) \equiv f^{\uparrow}(x,\mu^2) - f^{\downarrow}(x,\mu^2)
\,\mbox{,}
\ \ \ \ \ \ \ \ \ \
f=u,\bar{u},d,\bar{d},s,\bar{s},g
\,\mbox{,}
\label{eq:polPDFs}
\end{equation}
defined as the momentum densities of partons with spin 
aligned along ($^\uparrow$) or opposite ($^\downarrow$) the polarization 
direction of the parent nucleon. Polarized PDFs encode the spin structure 
of the nucleon, since these are related to the first two terms in 
Eq.~\eqref{eq:helicitysumrule}:
\begin{equation}
\Delta\Sigma(\mu^2)
=
\sum_{q=u,d,s}\int_0^1 dx \left[\Delta q(x, \mu^2) + \Delta\bar{q}(x, \mu^2)\right]
\ \  
\Delta G(\mu^2)
=
\int_0^1 dx \Delta g(x,\mu^2)
\,\mbox{.}
\label{eq:moments}
\end{equation}

The dependence of the PDFs on $x$, the momentum fraction of the proton carried 
by the parton, is genuinely nonperturbative, and, as such, must be inferred 
from data. These are usually supplemented with some theoretical constraints in 
order to achieve a meaningful PDF determination. First, PDFs must lead to 
positive cross sections: at leading order (LO), this implies that polarized 
PDFs are bounded by their unpolarized counterparts\footnote{Beyond LO, more 
complicate relations hold, but they have negligible impact on  
PDFs~\cite{Altarelli:1998gn}.}, $|\Delta f(x,\mu^2)|\leq f(x,\mu^2)$.
Second, polarized PDFs must be integrable: this corresponds to the assumption 
that the nucleon matrix element of the axial current for each flavor is finite.
Third, it follows from SU(2) and SU(3) flavor symmetry that 
the first moments of the nonsinglet $\mathcal{C}$-even PDF combinations,
$\Delta T_3=\Delta u^+ -\Delta d^+$ and 
$\Delta T_8 = \Delta u^+ +\Delta d^+ -2\Delta s^+$ 
(where $\Delta q^+=\Delta q+\Delta\bar{q}$, $q=u,d,s$), are 
related to the baryon octet $\beta$-decay constants, whose values are 
well measured~\cite{Agashe:2014kda}:
\begin{equation}
 a_3=\int_0^1 dx \Delta T_3 = 1.2701 \pm 0.0025
 \ \ \ \ \ \ \ \ \ \
 a_8=\int_0^1 dx \Delta T_8 = 0.585  \pm 0.025
 \,\mbox{.}
\label{eq:decayconst}
\end{equation}

The dependence of the PDFs on $\mu$ is perturbative. This obeys 
Dokshitzer-Gribov-Lipatov-Altarelli-Parisi (DGLAP) evolution 
equations~\cite{Altarelli:1977zs}, and has been computed up to 
next-to-next-to-leading order (NNLO).\footnote{In the polarized case,
the computation has been completed only very recently~\cite{Moch:2014sna}.} 
In principle, a determination of polarized PDFs at NNLO is then possible, 
even though based on inclusive DIS data alone: indeed, among the observables 
relevant for a determination of polarized PDFs, coefficient functions are known 
at NNLO only for the polarized DIS structure function 
$g_1$~\cite{Zijlstra:1993sh}. 
In practice, the impact of NNLO corrections is smaller than the current 
experimental uncertainties on data, hence moving from next-to-leading order 
(NLO) to NNLO determinations of polarized PDFs is not convenient so far.

Several determinations of polarized PDFs of the proton (up to NLO and mostly 
in the $\overline{\rm MS}$ factorization scheme) are presently available, 
aiming at unveiling how much large and uncertain are $\Delta\Sigma$ and 
$\Delta G$, Eq.~\eqref{eq:moments}. Above all, they differ among each others 
for the procedure used to determine PDFs from data and for the included data 
sets (for details, see {\it e.g.} Chap.~3 in 
Ref.~\refcite{Nocera:2014vla}). Both are a source of uncertainty, but the 
former may be elusive: in this respect, the {\tt NNPDF} collaboration has 
developed a methodology to reduce and keep it under control as much as possible.
This is based on cutting-edge statistical tools, including Monte Carlo sampling
for error propagation, neural networks for PDF parametrization, and closure 
tests for explicit characterization of procedural 
uncertainties~\cite{Ball:2014uwa}.

The bulk of the experimental information on polarized PDFs comes from 
neutral-current inclusive and semi-inclusive Deep-Inelastic Scattering 
(DIS and SIDIS) with charged lepton beams and nuclear targets. Because of the
way the corresponding observables factorize, inclusive DIS data constrain the 
total quark combinations $\Delta q^+$, 
while SIDIS data, with identified pions or kaons in the final state, 
constrain individual quark and antiquark flavors. In principle, both DIS and 
SIDIS data would constrain the gluon distribution $\Delta g$ via scaling 
violations, but in practice their effect is rather weak because of the small 
$Q^2$ range covered. 

Beside DIS and SIDIS fixed-target data, a significant amount of data from
longitudinally polarized proton-proton ($pp$) collisions at the Relativistic 
Heavy Ion Collider (RHIC) have become available 
recently~\cite{Aschenauer:2015eha}, though in a limited range of momentum 
fractions, $0.05\lesssim x \lesssim 0.4$.
On the one hand, longitudinal (parity-violating) single-spin and 
(parity-conserving) double-spin asymmetries for $W^\pm$ boson production are 
sensitive to the flavor decomposition of polarized quark and antiquark 
distributions, because of the chiral nature of the weak 
interactions~\cite{Bourrely:1993dd}. On the other hand, 
double-spin asymmetries for jet and $\pi^0$ 
production are directly sensitive to the gluon polarization in 
the proton, because of the dominance of gluon-gluon and quark-gluon initiated 
subprocesses in the kinematic range accessed by RHIC~\cite{Bourrely:1990pz}.

Motivated by the interest in studying the effects of this piece of experimental
information, two new global analyses of polarized PDFs have been carried out in
2014, {\tt DSSV14}~\cite{deFlorian:2014yva} and 
{\tt NNPDFpol1.1}~\cite{Nocera:2014gqa}. 
These upgrade the corresponding previous analyses, 
{\tt DSSV08}~\cite{deFlorian:2008mr} and 
{\tt NNPDFpol1.0}~\cite{Ball:2013lla}, with data respectively on double-spin 
asymmetries for inclusive jet production~\cite{Adamczyk:2014ozi} 
and $\pi^0$ production~\cite{Adare:2014hsq}\footnote{Preliminary RHIC results 
included in Ref.~\refcite{deFlorian:2008mr} have been replaced in
Ref.~\refcite{deFlorian:2014yva} with final results.}, 
and on double-spin asymmetries for high-$p_T$ inclusive jet 
production~\cite{Adamczyk:2014ozi,Adamczyk:2012qj,Adare:2010cc} and single-spin
asymmetries for $W^\pm$ production~\cite{Adamczyk:2014xyw}.
Some new data by the COMPASS experiment have also been included in 
{\tt DSSV14} and {\tt NNPDFpol1.1}, respectively new DIS and SIDIS 
data~\cite{Alekseev:2010hc,Alekseev:2010ub} and open-charm leptoproduction 
data~\cite{Adolph:2012ca}. The new data have been included in {\tt NNPDFpol1.1}
by means of Bayesian reweighting~\cite{Ball:2010gb},
and in {\tt DSSV14} by means of a full refit.  

Overall, both {\tt DSSV14} and {\tt NNPDFpol1.1} PDF determinations are 
state-of-the-art in the inclusion of the available experimental information. 
The data sets in the two analyses differ between each other only for 
fixed-target SIDIS and RHIC $\pi^0$ production measurements, included in 
{\tt DSSV14}, but not in {\tt NNPDFpol1.1}. The information brought in by 
these data is complementary to that provided by RHIC $W^\pm$ production and 
inclusive jet production data respectively, but this is less 
constraining~\cite{Nocera:2014gqa}. These data were not included in the 
{\tt NNPDFpol1.1} analysis because fragmentation functions (FFs) enter the 
factorized expression of the corresponding observables: since FFs are 
nonperturbative objects on the same footing as PDFs, they are likely to 
introduce an additional source of bias in the PDF determination. The
{\tt NNPDF} methodology aims at reducing this bias as much as possible, hence 
the inclusion of these data would require the consistent determination 
of FFs within the {\tt NNPDF} methodology, which is under consideration,
though not yet available~\cite{Bertone:2015cwa}. 

The effect of RHIC data on the polarized PDFs of the proton is twofold.  
\begin{itemlist}
\item The 2012 STAR data sets on $W$ production~\cite{Adamczyk:2014xyw}, 
included in {\tt NNPDFpol1.1}, provide evidence of a positive 
$\Delta\bar{u}$ distribution 
and a negative $\Delta\bar{d}$ distribution, with 
$|\Delta\bar{d}|>|\Delta\bar{u}|$~\cite{Nocera:2014gqa}. 
The size of this flavor symmetry breaking for polarized sea quarks is 
quantified by the asymmetry $\Delta\bar{u}-\Delta\bar{d}$, which,
in the {\tt NNPDFpol1.1} analysis, turned out to be roughly as large as its 
unpolarized counterpart (in absolute value), 
though much more uncertain~\cite{Nocera:2014rea}. Even within this uncertainty,
polarized and unpolarized light sea quark asymmetries show opposite sings,
with the polarized being definitely positive.  
This result starts to discriminate between different 
models of nucleon structure, see left panel of Fig.~\ref{fig:RHICpdfs}: 
specifically, some meson-cloud (MC) models are disfavored, while a more 
accurate experimental information is needed to establish whether 
chiral quark-soliton (CQS), Pauli-blocking (PB) or statistical (ST)
models are preferred (these models are described in 
Ref.~\refcite{Chang:2014jba}).
\begin{figure}[t]
\centerline
{
\includegraphics[scale=0.29]{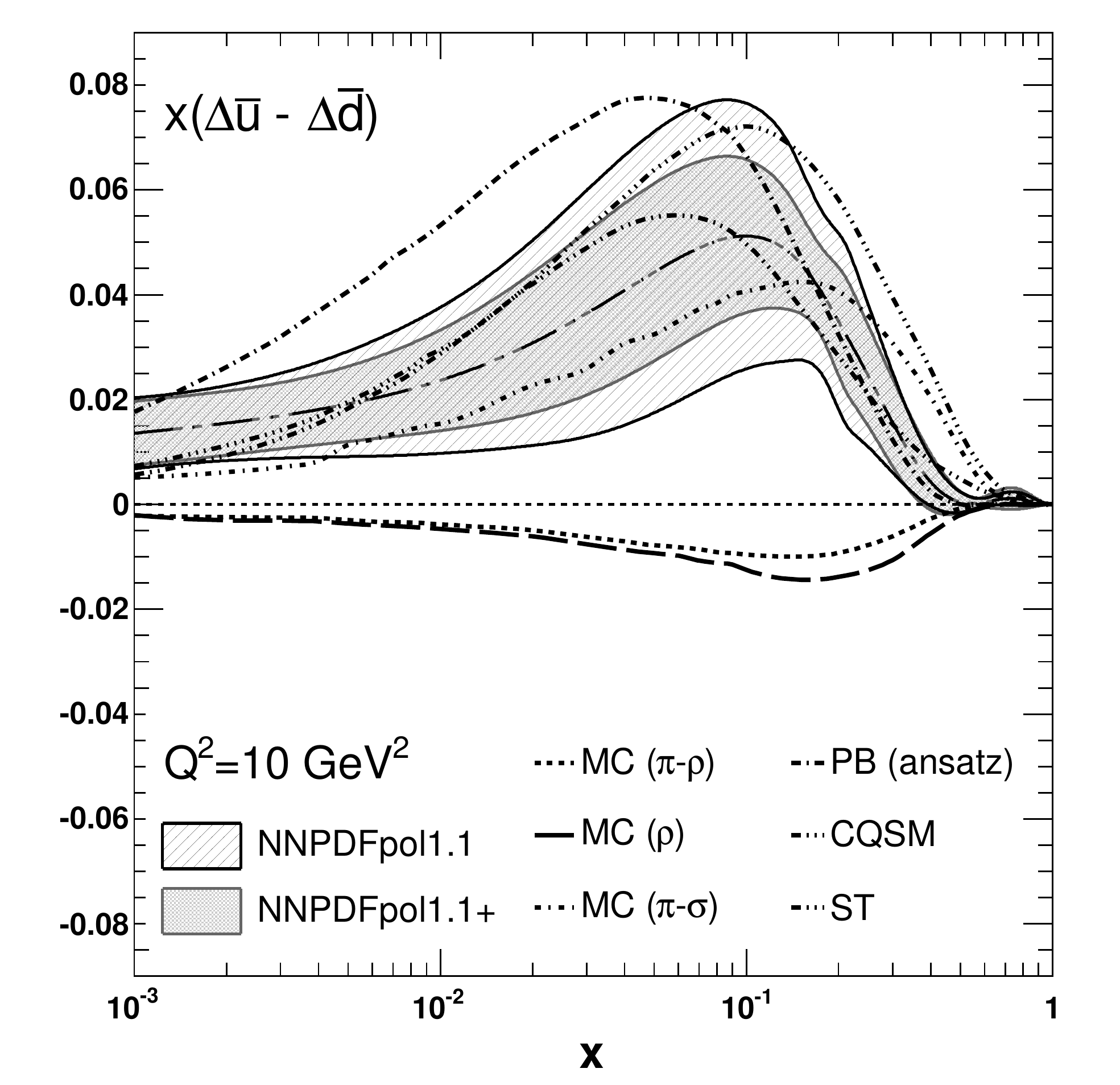}
\includegraphics[scale=0.29]{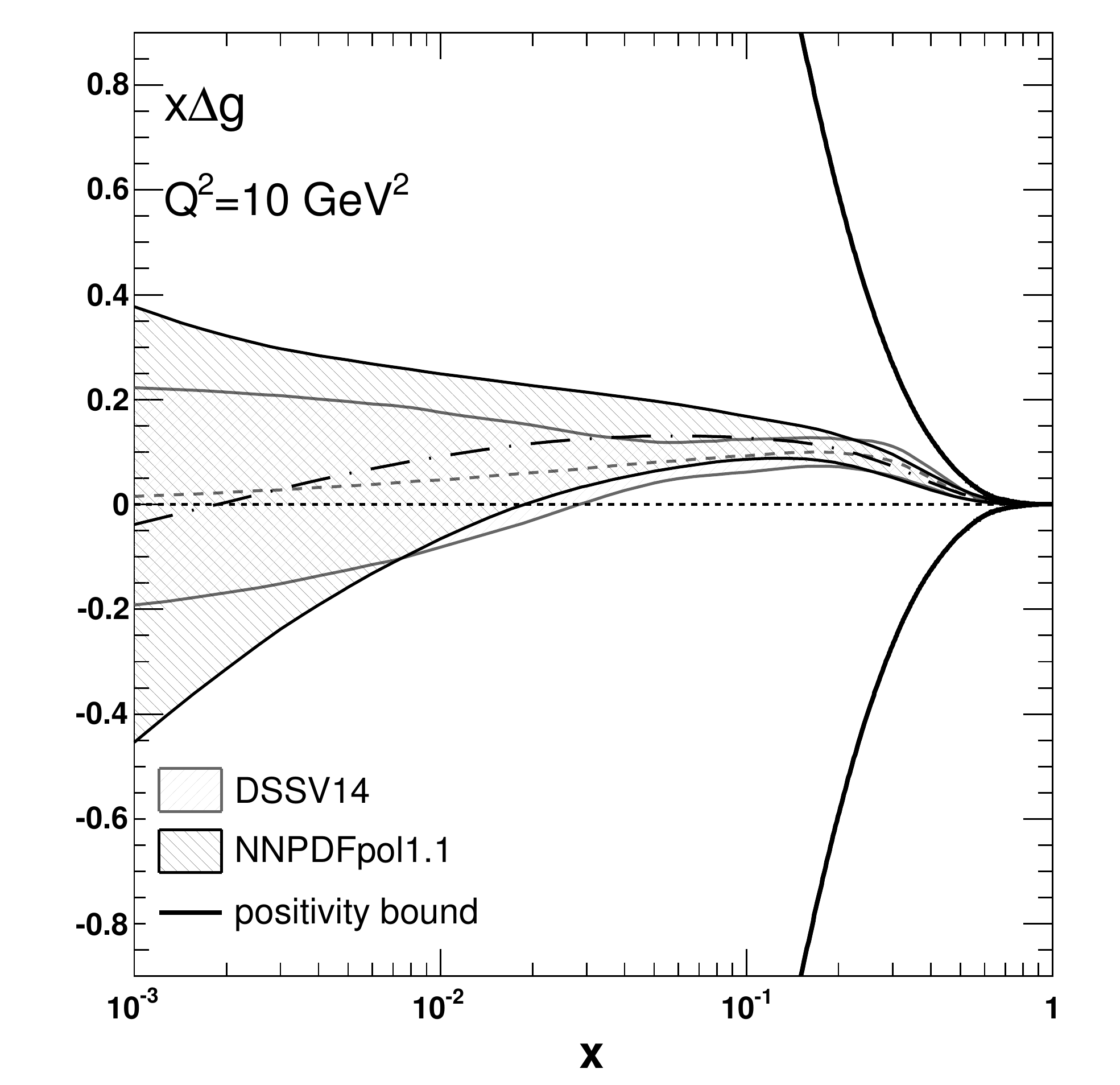}\\
}
\vspace*{8pt}
\caption{(Left) The polarized light sea quark asymmetry 
$x(\Delta\bar{u}-\Delta\bar{d})$ from the {\tt NNPDFpol1.1} and 
{\tt NNPDFpol1.1+} (supplemented with pseudodata for $W^\pm$ production at RHIC)
PDF sets at $Q^2=10$ GeV$^2$, compared to expectations from 
various models of nucleon structure~\cite{Chang:2014jba}. 
(Right) The polarized gluon $x\Delta g$ from the 
{\tt DSSV14} and {\tt NNPDFpol1.1} PDF sets at $Q^2=10$ GeV$^2$.}
\label{fig:RHICpdfs}
\end{figure}
\item The 2009 STAR and PHENIX data sets on jet and $\pi^0$ 
production~\cite{Adamczyk:2014ozi,Adare:2014hsq}, included in both {\tt DSSV14}
and {\tt NNPDFpol1.1}, provide first evidence
of a sizable, positive gluon polarization in the proton. 
A comparison of the gluon PDF in the two PDF sets is displayed in 
Fig.~\ref{fig:RHICpdfs} (right panel). Comparable results, both central values 
and uncertainties, are found in the $x$ region covered by RHIC data. 
The agreement between the two analyses is optimal in the
range $0.08\leq x \leq 0.2$, where the dominant experimental information comes
from jet data; a slightly smaller central value is found in the {\tt DSSV14} 
analysis, in comparison to the {\tt NNPDFpol1.1}, in the range 
$0.05\leq x \leq 0.08$, where the dominant experimental information comes from 
$\pi^0$ production data. Indeed, these are included in {\tt DSSV14} but are not
in {\tt NNPDFpol1.1}. Nevertheless, best fits lie well within each other error
bands, though {\tt NNPDF} uncertainties tend to be larger than {\tt DSSV14}
uncertainties outside the region covered by RHIC data.
Very well compatible values of the integral of $\Delta g$, 
Eq.~\eqref{eq:moments}, truncated over the interval $0.05\leq x \leq 1$, are 
found: at $Q^2=10$ GeV$^2$, this is $0.20^{+0.06}_{-0.07}$ for 
{\tt DSSV14}~\cite{deFlorian:2014yva}, and $0.23\pm 0.06$ for 
{\tt NNPDFpol1.1}~\cite{Nocera:2014gqa}.
\end{itemlist}

\begin{figure}[pt]
\centerline
{
\includegraphics[scale=0.29]{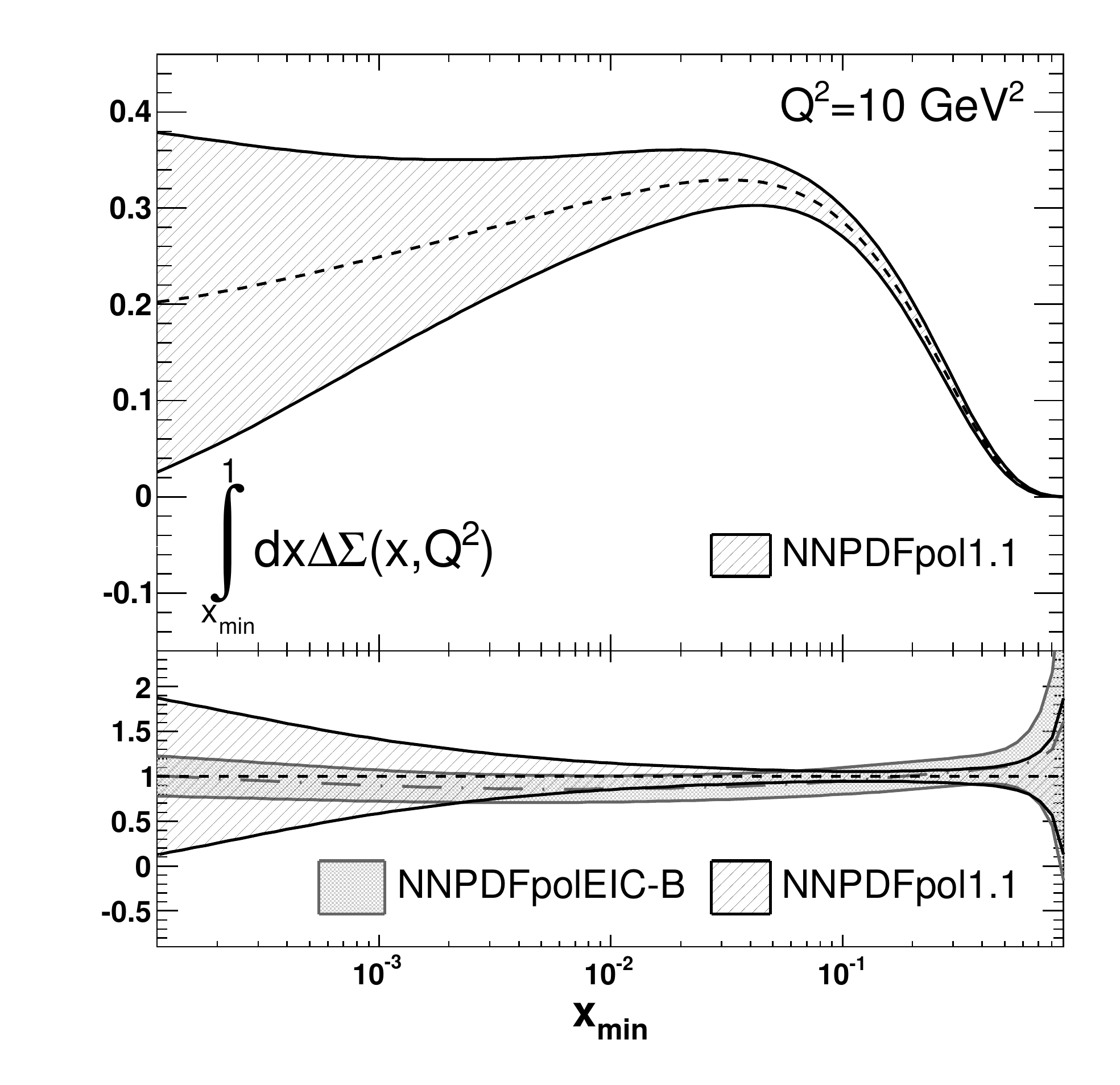}
\includegraphics[scale=0.29]{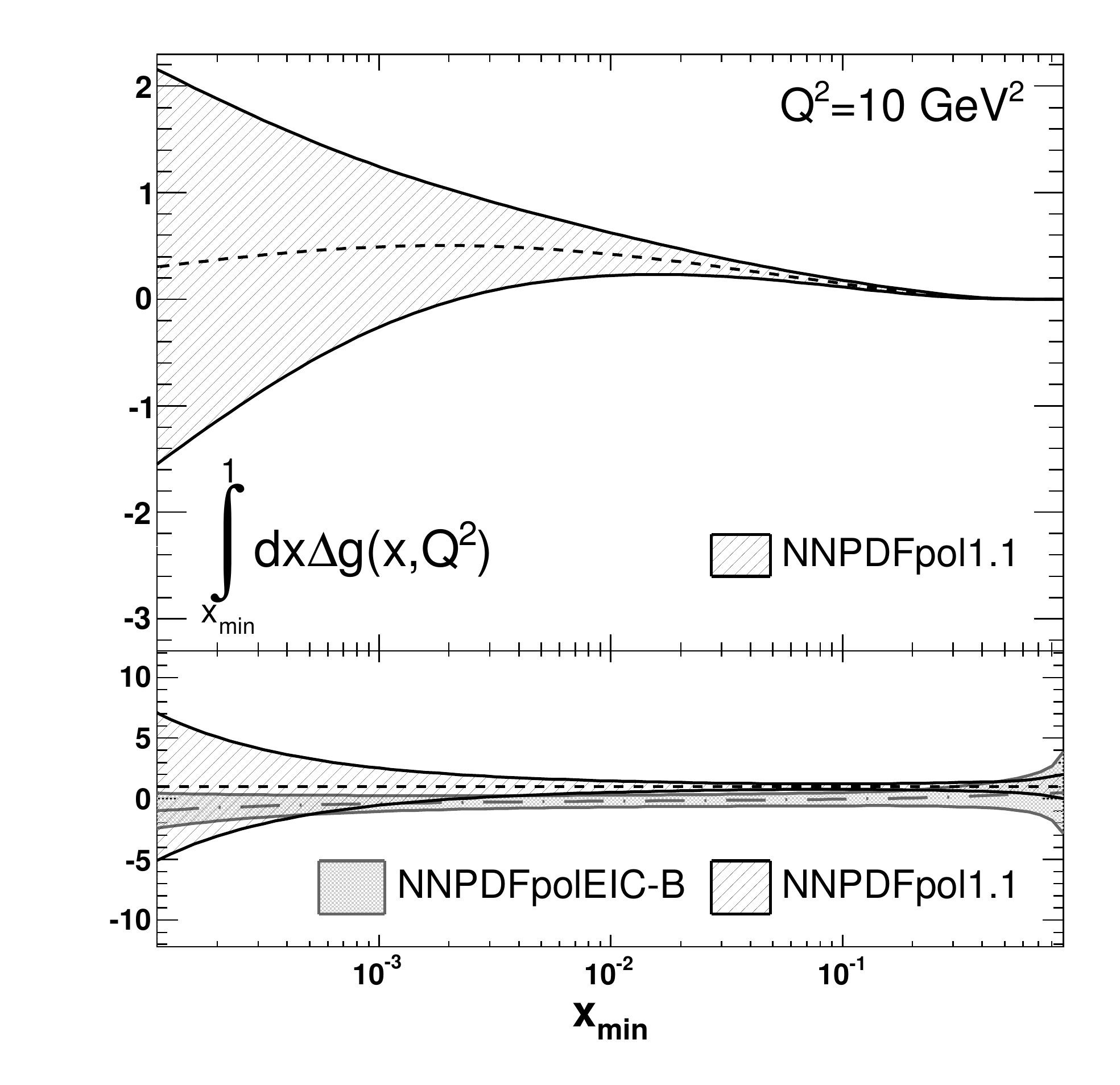}
}
\vspace*{8pt}
\caption{The running integral $\int_{x_{\rm min}}^1 dx \Delta f(x,Q^2)$, 
$f=\Sigma, g$, from {\tt NNPDFpol1.1}~\cite{Nocera:2014gqa} at $Q^2=10$ GeV$^2$ 
(upper panels), and their ratio to {\tt NNPDFpolEIC-B}~\cite{Ball:2013tyh}
based on EIC pseudodata (lower panels).}
\label{fig:runningmoments}
\end{figure}

Despite the achievements described above, the lack of experimental data 
over a wide range of $x$ and $Q^2$ values seriously limits the accuracy with 
which polarized PDFs can be determined. Several issues in our knowledge of the 
nucleon longitudinal spin structure are hence left completely open, 
as summarized below.  
\begin{itemlist}
 \item The size of the contribution of quarks, antiquarks and gluons to the 
nucleon spin, as quantified by their first moments, Eq.~\eqref{eq:moments}, 
is affected by large uncertainties. These come predominantly from the 
extrapolation into the small-$x$ region ($x\lesssim 10^{-3}$), not covered by 
experimental data. In order to illustrate the situation, the {\it running} 
integrals of singlet and gluon distributions ({\it i.e.} the quantities in 
Eq.~\eqref{eq:moments} evaluated as a function of the lower limit of 
integration $x_{\rm min}$) are displayed in Fig.~\ref{fig:runningmoments} 
at $Q^2=10$ GeV$^2$. It is apparent that, as $x_{\rm min}$ decreases, the 
uncertainty on the integrals increases up to a size that prevents from any
firm conclusion on their contribution to Eq.~\eqref{eq:helicitysumrule}. 
 \item A precision test of the Bjorken sum rule~\cite{Bjorken:1966jh} is 
presently not achievable within the accuracy of available data: indeed, 
a largely uncertain, and potentially substantial, contribution to it may arise 
in the small-$x$ region, $x\lesssim 10^{-3}$~\cite{Ball:2013lla}. As a 
consequence, a determination of $\alpha_s$ from the Bjorken sum rule 
is not competitive.
 \item Fairly significant violations of SU(3) symmetry are advocated
in the literature (see {\it e.g.} Ref.~\refcite{Cabibbo:2003cu} for a review). 
In this case, an uncertainty on the octet axial charge, 
larger up to $30\%$ than its experimental value, Eq.~\eqref{eq:decayconst}, 
is found~\cite{FloresMendieta:1998ii}. It was shown~\cite{Ball:2013lla} that
a more conservative estimate of the uncertainty on $a_8$ has a limited 
impact on the behavior of polarized PDFs in a global analysis. 
Nevertheless, if the octet sum rule in Eq.~\eqref{eq:decayconst} were not 
imposed at all, data alone would not be sufficiently accurate 
either to constrain $\Delta T_8$ or to test potential SU(3) violations. 
 \item Inclusive DIS data, together with nonsinglet axial 
couplings, Eq.~\eqref{eq:decayconst}, and kaon-tagged SIDIS data 
provide the sole constraint on the total strange distribution $\Delta s^+$. 
No experimental information is available on individual $\Delta s$ and 
$\Delta\bar{s}$ PDFs, and $\Delta s=\Delta\bar{s}$ is often assumed for
phenomenological purposes. Mutually consistent, sizable, negative values of 
the first moment of $\Delta s^+$ are found in both 
{\tt DSSV}~\cite{deFlorian:2008mr,deFlorian:2014yva} 
and {\tt NNPDF}~\cite{Ball:2013lla,Nocera:2014gqa} analyses, though they arise
from rather different shapes of $\Delta s^+$. This discrepancy may considerably
depend on the kaon FF used to analyze SIDIS data~\cite{Leader:2011tm}, 
which are included in the {\tt DSSV} determinations and are not in the 
{\tt NNPDF} determinations. Further measurements of spin-dependent kaon
production cross section in SIDIS and an improved determination of the kaon FF
are needed to make any definitive conclusion. 
 \item Several models of nucleon structure have been developed, aiming at 
predicting the polarized PDF behavior at small and large $x$. In order to test
whether data favor or unfavor them, computations of spin-dependent observables 
based on these models should be compared to the corresponding expectations 
based on a global analysis of polarized PDFs. Unfortunately, only a limited 
number of models are clearly incompatible with the predictions based on 
{\tt DSSV14} and {\tt NNPDFpol1.1}~\cite{Nocera:2014uea}. More abundant 
and more accurate data at small and large $x$ would be needed to discriminate 
among the other models.
\end{itemlist}

An intense experimental campaign is being devised to address these issues in 
the future. Some ongoing, planned and proposed prospects are discussed in the 
sequel, in order of increasing impact on the proton longitudinal spin 
structure.
\begin{itemlist}
 \item The COMPASS experiment at CERN is about completing its experimental 
program dedicated to the longitudinal spin structure of the 
nucleon. It will provide additional high-precision 
results for the inclusive DIS structure function of the proton 
$g_1^p$~\cite{Wilfert:2014tea}. However, these are expected to be of limited 
impact in narrowing the small-$x$ extrapolation uncertainty on the full first 
moments, Eq.~\eqref{eq:moments}, since they will extend only down to 
$x\sim 0.004$.
 \item The CLAS and Hall-A experiments at JLAB have recently presented a large
amount of high-precision data respectively on the ratio of polarized to 
unpolarized proton and deuteron structure functions, 
$g_1^{p,d}/F_1^{p,d}$~\cite{Prok:2014ltt}, and on the proton and neutron spin
photoabsorption asymmetry, $A_1^{p,n}$~\cite{Parno:2014xzb}. Since these data are
taken in a kinematic region (large $x$, small $Q^2$) where dynamic higher-twist
contributions to the Wilson expansion of $g_1$ and resummation effects may be 
relevant~\cite{Jimenez-Delgado:2013boa,Anderle:2013lka}, their inclusion in a 
global analysis of polarized PDFs is highly nontrivial, and not yet performed.
The kinematic reach of the existing JLAB data is expected to extend to twice
smaller $x$ as well as to larger $x$ values after the $12$ GeV electron beam 
energy upgrade~\cite{Dudek:2012vr}: large luminosity and high resolution 
available will allow for a substantial reduction of PDF uncertainties in the 
medium-to-large $x$ region.
 \item The STAR and PHENIX experiments at RHIC are expected to provide 
high-impact results in the near term~\cite{Aschenauer:2015eha}. 
As for $W^\pm$ production, much smaller uncertainties on the single-spin 
asymmetry will be reached thanks to the substantially increased statistics 
from run-2013. 
Supplementing {\tt NNPDFpol1.1} with $W^\pm$ production pseudodata based on 
projected uncertainties will reduce the uncertainty on the polarized light sea 
quark asymmetry by a factor two, see Fig.~\ref{fig:RHICpdfs}
(the PDF determination including pseudodata is labeled {\tt NNPDFpol1.1+}).
As for $\pi^0$ and jet production, future RHIC measurements up to run-2015 
at center-of-mass energy $\sqrt{s}=510$ GeV will be sensitive to the polarized 
gluon PDF down to $x\sim 10^{-3}$. Inclusion of $\pi^0$ and jet production 
pseudodata based on projected uncertainties in the {\tt DSSV14} analysis will 
reduce the uncertainty on $\Delta g$ by a factor two~\cite{Aschenauer:2015eha}.
Complementary information on $\Delta g$ will be provided by correlation 
measurements, {\it e.g.} in di-jet and di-hadron production processes.
 \item A future high-energy, polarized Electron-Ion Collider 
(EIC)~\cite{Accardi:2012qut} will likely be the only facility to address all
the above questions with the highest precision. 
The extension of the kinematic reach down to $x\sim 10^{-4}$ and up to
$Q^2=10^4$ GeV$^2$ will allow for an accurate determination of $\Delta g$
via scaling violations in inclusive DIS, of $\Delta\bar{u}$ and 
$\Delta\bar{d}$ via inclusive DIS at high $Q^2$ mediated by electroweak bosons,
and of $\Delta s$ via kaon-tagged SIDIS. Using simulated pseudodata at the 
eRHIC realization of an EIC~\cite{Aschenauer:2014cki}, the impact of some of 
these measurements has been recently 
studied~\cite{Aschenauer:2012ve,Ball:2013tyh}. 
It was found that the running integrals $\Delta\Sigma$ and $\Delta G$
will be determined with an accuracy of about respectively $\pm 20\%$ and
$\pm 10\%$, see Fig.~\ref{fig:runningmoments}. Would the EIC data 
confirm the {\tt DSSV14} or {\tt NNPDFpol1.1} best fit behaviors of 
$\Delta\Sigma$ and $\Delta G$, see Fig.~\ref{fig:runningmoments}, only a small 
fraction of the proton spin is expected to come from orbital angular momentum,
$\mathcal{L}_q+\mathcal{L}_g$ in Eq.~\eqref{eq:helicitysumrule}, 
at $Q^2=10$ GeV$^2$.
\end{itemlist}

A complete understanding of the origin of the proton spin is still lacking.
Brand new facilities, such as the EIC, will be eventually required to elucidate
how this is built up from the interplay between the intrinsic properties and 
interactions of quarks and gluons. Exploring the proton spin may lead to
unexpected discoveries in the future, like it happened in the past for the  
evidence of nucleon substructure in the measurement of the proton anomalous 
magnetic moment; ultimately, this may shed light on confinement, 
an essential step towards our understanding of QCD.

\bibliographystyle{ws-ijmpcs}
\bibliography{S2_Nocera.bib}

\begin{thebibliography}{10}

\bibitem{Bass:2007zzb}
S.~D. Bass, {\em The Spin structure of the proton} (World Scientific, 2007).

\bibitem{Jaffe:1989jz}
R.~Jaffe and A.~Manohar, {\em Nucl.Phys.} {\bf B337}, 509  (1990).

\bibitem{Collins:2011zzd}
J.~Collins, {\em Foundations of perturbative QCD} (Cambridge, 2011).

\bibitem{Leader:2013jra}
E.~Leader and C.~Lorc\'{e}, {\em Phys.Rept.} {\bf 541}, 163  (2013).

\bibitem{Aidala:2012mv}
C.~A. Aidala, S.~D. Bass, D.~Hasch and G.~K. Mallot, {\em Rev.Mod.Phys.} {\bf
  85}, 655  (2013).

\bibitem{Collins:1989gx}
J.~C. Collins, D.~E. Soper and G.~F. Sterman, {\em Adv.Ser.Direct.High Energy
  Phys.} {\bf 5}, 1  (1988), J.~C. Collins, {\it Nucl.Phys.} {\bf B394}, 169
  (1993).

\bibitem{Altarelli:1998gn}
G.~Altarelli, S.~Forte and G.~Ridolfi, {\em Nucl.Phys.} {\bf B534}, 277
  (1998).

\bibitem{Agashe:2014kda}
K.~Olive {\em et~al.}, {\em Chin.Phys.} {\bf C38}, 090001  (2014).

\bibitem{Altarelli:1977zs}
G.~Altarelli and G.~Parisi, {\em Nucl.Phys.} {\bf B126}, 298  (1977).

\bibitem{Moch:2014sna}
S.~Moch, J.~Vermaseren and A.~Vogt, {\em Nucl.Phys.} {\bf B889}, 351  (2014).

\bibitem{Zijlstra:1993sh}
E.~Zijlstra and W.~van Neerven, {\em Nucl.Phys.} {\bf B417}, 61  (1994).

\bibitem{Nocera:2014vla}
E.~R. Nocera  (2014), PhD thesis, arXiv:1403.0440.

\bibitem{Ball:2014uwa}
R.~D. Ball {\em et~al.}  (2014), arXiv:1410.8849.

\bibitem{Aschenauer:2015eha}
E.~Aschenauer {\em et~al.}  (2015), arXiv:1501.01220.

\bibitem{Bourrely:1993dd}
C.~Bourrely and J.~Soffer, {\em Phys.Lett.} {\bf B314}, 132  (1993).

\bibitem{Bourrely:1990pz}
C.~Bourrely, J.~Guillet and J.~Soffer, {\em Nucl.Phys.} {\bf B361}, 72  (1991).

\bibitem{deFlorian:2014yva}
D.~de~Florian {\em et~al.}, {\em Phys.Rev.Lett.} {\bf 113}, 012001  (2014).

\bibitem{Nocera:2014gqa}
E.~R. Nocera {\em et~al.}, {\em Nucl.Phys.} {\bf B887}, 276  (2014).

\bibitem{deFlorian:2008mr}
D.~de~Florian {\em et~al.}, {\em Phys.Rev.Lett.} {\bf 101}, 072001  (2008),
  {\it Phys.Rev.} {\bf D80}, 034030 (2009).

\bibitem{Ball:2013lla}
R.~D. Ball {\em et~al.}, {\em Nucl.Phys.} {\bf B874}, 36  (2013).

\bibitem{Adamczyk:2014ozi}
L.~Adamczyk {\em et~al.}  (2014), arXiv:1405.5134.

\bibitem{Adare:2014hsq}
A.~Adare {\em et~al.}, {\em Phys.Rev.} {\bf D90}, 012007  (2014).

\bibitem{Adamczyk:2012qj}
L.~Adamczyk {\em et~al.}, {\em Phys.Rev.} {\bf D86}, 032006  (2012).

\bibitem{Adare:2010cc}
A.~Adare {\em et~al.}, {\em Phys.Rev.} {\bf D84}, 012006  (2011).

\bibitem{Adamczyk:2014xyw}
L.~Adamczyk {\em et~al.}, {\em Phys.Rev.Lett.} {\bf 113}, 072301  (2014).

\bibitem{Alekseev:2010hc}
M.~Alekseev {\em et~al.}, {\em Phys.Lett.} {\bf B690}, 466  (2010).

\bibitem{Alekseev:2010ub}
M.~Alekseev {\em et~al.}, {\em Phys.Lett.} {\bf B693}, 227  (2010).

\bibitem{Adolph:2012ca}
C.~Adolph {\em et~al.}, {\em Phys.Rev.} {\bf D87}, 052018  (2013).

\bibitem{Ball:2010gb}
R.~D. Ball {\em et~al.}, {\em Nucl.Phys.} {\bf B849}, 112  (2011),
  Erratum-ibid. {\bf B854}, 926 (2012), Erratum-ibid. {\bf B855}, 927 (2012),
  {\it Nucl.Phys.} {\bf B855}, 608 (2012).

\bibitem{Bertone:2015cwa}
V.~Bertone, S.~Carrazza and E.~R. Nocera  (2015), arXiv:1501.00494.

\bibitem{Nocera:2014rea}
E.~R. Nocera, {\em PoS} {\bf DIS2014}, 204  (2014).

\bibitem{Chang:2014jba}
W.-C. Chang and J.-C. Peng, {\em Prog.Part.Nucl.Phys.} {\bf 79}, 95  (2014).

\bibitem{Bjorken:1966jh}
J.~Bjorken, {\em Phys.Rev.} {\bf 148}, 1467  (1966), {\it Phys.Rev.} {\bf D1},
  1376 (1970).

\bibitem{Cabibbo:2003cu}
N.~Cabibbo, E.~C. Swallow and R.~Winston, {\em Ann.Rev.Nucl.Part.Sci.} {\bf
  53}, 39  (2003).

\bibitem{FloresMendieta:1998ii}
R.~Flores-Mendieta, E.~E. Jenkins and A.~V. Manohar, {\em Phys.Rev.} {\bf D58},
  094028  (1998).

\bibitem{Leader:2011tm}
E.~Leader, A.~V. Sidorov and D.~B. Stamenov, {\em Phys.Rev.} {\bf D84}, 014002
  (2011).

\bibitem{Nocera:2014uea}
E.~R. Nocera, {\em Phys.Lett.} {\bf B742}, 117  (2015).

\bibitem{Ball:2013tyh}
R.~D. Ball {\em et~al.}, {\em Phys.Lett.} {\bf B728}, 524  (2014).

\bibitem{Wilfert:2014tea}
M.~Wilfert, {\em PoS} {\bf DIS2014}, 206  (2014).

\bibitem{Prok:2014ltt}
Y.~Prok {\em et~al.}, {\em Phys.Rev.} {\bf C90}, 025212  (2014).

\bibitem{Parno:2014xzb}
D.~Parno {\em et~al.}  (2014), arXiv:1406.1207.

\bibitem{Jimenez-Delgado:2013boa}
P.~Jimenez-Delgado, A.~Accardi and W.~Melnitchouk, {\em Phys.Rev.} {\bf D89},
  034025  (2014).

\bibitem{Anderle:2013lka}
D.~P. Anderle, F.~Ringer and W.~Vogelsang, {\em Phys.Rev.} {\bf D87}, 094021
  (2013).

\bibitem{Dudek:2012vr}
J.~Dudek, R.~Ent, R.~Essig, K.~Kumar, C.~Meyer {\em et~al.}, {\em Eur.Phys.J.}
  {\bf A48}, 187  (2012).

\bibitem{Accardi:2012qut}
A.~Accardi {\em et~al.}  (2012), arXiv:1212.1701.

\bibitem{Aschenauer:2014cki}
E.~Aschenauer {\em et~al.}  (2014), arXiv:1409.1633.

\bibitem{Aschenauer:2012ve}
E.~Aschenauer {\em et~al.}, {\em Phys.Rev.} {\bf D86}, 054020  (2012), {\it
  Phys.Rev.} {\bf D88}, 114025 (2013).

\end{thebibliography}

\end{document}